\documentstyle[prl,aps]{revtex}
\input epsf

\tighten
\begin{document}

\def\be{\begin{equation}}
\def\ee{\end{equation}}
\def\ba{\begin{eqnarray}}
\def\ea{\end{eqnarray}}
\def\bq{\begin{quote}}
\def\eq{\end{quote}}
\def\PL{{ \it Phys. Lett.} }
\def\PRL{{\it Phys. Rev. Lett.} }
\def\NP{{\it Nucl. Phys.} }
\def\PR{{\it Phys. Rev.} }
\def\MPL{{\it Mod. Phys. Lett.} }
\def\IJMP{{\it Int. J. Mod .Phys.} }
\newcommand{\labell}[1]{\label{#1}\qquad_{#1}} 
\newcommand{\labels}[1]{\vskip-2ex$_{#1}$\label{#1}} 
\newcommand\gapp{\mathrel{\raise.3ex\hbox{$>$}\mkern-14mu
\lower0.6ex\hbox{$\sim$}}}
\newcommand\gsim{\gapp}
\newcommand\gtsim{\gapp}
\newcommand\lapp{\mathrel{\raise.3ex\hbox{$<$}\mkern-14mu
\lower0.6ex\hbox{$\sim$}}}
\newcommand\lsim{\lapp}
\newcommand\ltsim{\lapp}
\newcommand\M{{\cal M}}
\newcommand\order{{\cal O}}

\newcommand\extra{{\rm {extra}}}
\newcommand\FRW{{\rm {FRW}}}
\newcommand\brm{{\rm {b}}}
\newcommand\ord{{\rm {ord}}}
\newcommand\Pl{{\rm {pl}}}
\newcommand\Mpl{M_{\rm {pl}}}
\newcommand\mgap{m_{\rm {gap}}}
\newcommand\gB{g^{\left(\rm \small B\right)}}

\preprint{SU-GP-01/7-1\\ July 2001}

\draft
\title{Leptogenesis and Neutrinos at a TeV}

\author{Salah Nasri and Mark Trodden}

\address
{Department of Physics \\
Syracuse University \\
Syracuse, NY, 13244-1130 USA \\
}

\wideabs{

\maketitle

\begin{abstract}
\widetext We consider a model for leptogenesis in which all the relevant physics is accessible
in collider experiments. This requires a moderate extension of the standard model
which generates small neutrino masses via TeV scale right-handed neutrinos. The necessary
Sakharov criteria are satisfied in such a way that all existing experimental constraints are
satisfied. We demonstrate how the requisite baryon to
entropy ratio is generated, and discuss some possible experimental signatures
of the model.
\end{abstract}
\pacs{PACS:11.30.Fs, 12.60.-i \hfill SU-GP-01/7-1 \\
13.35.Hb, 98.80.Cq \hfill SU-4252-742}
}

\narrowtext
The discovery of neutrino masses and mixing, reported by the Super-Kamiokande \cite{superk}
and SNO \cite{SNO}
experiments, is an exciting example of the symbiosis
between particle physics and cosmology. One interesting consequence of this discovery for cosmology
is the possibility of generating the baryon asymmetry of the universe through leptogenesis via
the decays of right-handed neutrinos (for reviews see \cite{Riotto:1999yt}). However, one drawback of
this approach is its inaccessibility to independent
collider tests. This is because, in typical models, the masses of
the right-handed neutrinos must be of order $10^{10}$ GeV in order that realistic Majorana masses for left-handed
neutrinos be generated by the seesaw mechanism \cite{seesaw}. In this letter we describe a realistic model in
which leptogenesis is efficient, and the required right-handed neutrinos need only have TeV-scale masses. This
model is therefore a candidate for generating the baryon asymmetry of the universe (BAU), which may be
directly tested in upcoming experiments.

Leptogenesis as a precursor to baryogenesis works because
the $SU(2)_L$ sphaleron configurations which violate both baryon number (B) and lepton number (L),
nevertheless conserve the combination $B - L$. Thus, the baryon asymmetry may be generated from
an existing lepton
asymmetry. Indeed, an analysis \cite{harvey}, (see also \cite{Luty}) of the chemical potentials in the standard model yields a relation
between the baryon asymmetry
\begin{equation}
\eta_B =\frac{n_b - n_{\overline b}}{n_s} \ ,
\end{equation}
where $n_{b({\bar b})}$ is the number density of (anti)baryons and $n_s$ is the entropy density,
and the corresponding B-L asymmetry, $\eta_{B-L}$ given by
\begin{equation}
\eta_B = \frac{28}{79}\eta_{(B - L)_P} \ ,
\end{equation}
where $(B- L)_P$ is the primordial lepton asymmetry generated above the electroweak scale. It is the
baryon asymmetry that is constrained by primordial nucleosynthesis to be $\eta_B \sim 10^{-10}$.

One important
way in which leptons may be produced is by the out of equilibrium decay of heavy right handed Majorana
neutrinos $N_R$. Once this asymmetry is produced it is then partially converted to a baryonic one.

In order that such an asymmetry survive, it is essential that $B-L$ violating processes be out of
equilibrium after the leptons are produced. This yields a constraint on the masses of the right
handed Majorana neutrinos, and hence on the left-handed neutrino masses.
Consider the $\Delta L = 2$ process
\begin{equation}
\nu_L + \nu_L \longrightarrow \Phi_{SM} + \Phi_{SM} \ ,
\end{equation}
where $\nu_L$ are left-handed neutrinos and $\Phi_{SM}$ is the standard model Higgs field,
occurring at the temperature $T\approx 100$ GeV.
By imposing the out of equilibrium condition we
obtain the constraint \cite{yanagida}
\begin{equation}
m_\nu < 50 keV \ .
\end{equation}
A stronger bound is obtained by requiring that the rate of the lepton violating interactions, for example
\begin{equation}
W^{\pm} + W^{\pm} \longrightarrow e^{\pm} + e^{\pm} \ ,
\end{equation}
mediated by virtual left handed Majorana neutrino exchange, be smaller that the
expansion rate of the universe at temperature $T = M_W$. This yields the constraint $m_\nu <20$ keV
\cite{sarkar}.

Note that this constraint is independent of the existence of right handed neutrinos or triplet higgs fields.
It is a result of the effective interaction after integrating out the heavy fields, and it applies
to every element of the mass matrix.

\begin{figure}[h]
\epsfxsize = 0.85 \hsize \epsfbox{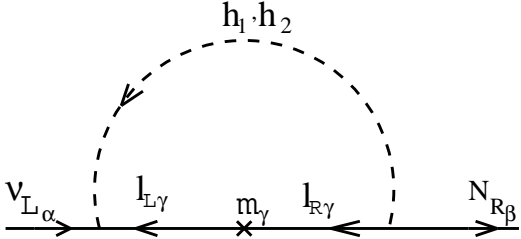}
\caption{\label{figure1}
A one-loop diagram generating Dirac masses for the neutrinos.
}
\end{figure}

In this paper, we discuss the possibility of generating the baryon asymmetry of the universe (BAU) via
leptogenesis at the TeV scale. We utilize a recently proposed model \cite{Nasri:2001ax} for neutrino
masses, in which the right handed Majorana masses are of order $10$ TeV. We extend the standard
model by adding
three right handed neutrinos $N_{R_{\alpha}}$ and two new charged higgs fields $S_1^{+}$
and $S_2^{+}$, which are singlets under $SU(2)_L$. The interaction of $S_1$ and $S_2$ with the standard
model fermions is given by:
\begin{equation}
{\cal L}_{int} = f_{\alpha\beta}L_{\alpha}^TCi\tau_2L_{\beta}S_1^{+} + g_{\alpha\beta}
l_{R_{\alpha}}^TCN_{R_{\beta}}S_2^{+} + {\rm h.c} \ ,
\label{interaction}
\end{equation}
where $\alpha$, $\beta$ denote generation indices. Here $L$ is the left-handed lepton
doublet, $l_{R_{\alpha}}$ are the right-handed charged leptons and
$C$ is the charge-conjugation matrix. The Yukawa couplings $f_{\alpha\beta}$ are antisymmetric
in $\alpha$ and $\beta$, while $g_{\alpha\beta}$ are arbitrary. We will demand that both $S_1^{+}$ and
$S_2^{+}$ carry lepton number $L = -2$ so that the above interactions
conserve lepton number.
The standard
model higgs doublet $\Phi_{SM}$ couples to the right handed neutrino and lepton
doublets, which leads to a tree level neutrino Dirac mass. In this case
unnatural fine tuning is needed. To forbid such coupling we impose a
discrete symmetry $Z_2$ acting as
\begin{eqnarray}
(L_\alpha, \Phi_{SM}, S_1^{+}) & \longrightarrow & (L_\alpha,\Phi_{SM},
S_1^{+}) \nonumber \\
(N_{R_{\alpha}}, S_2^{+}) & \longrightarrow & - (N_{R_{\alpha}}, S_2^{+}) \ .
\end{eqnarray}
If this $Z_2$ symmetry is exact, the neutrinos will be massless.  We therefore introduce the
soft symmetry-breaking term
\begin{equation}
\delta V = \kappa S_1S_2^{+} + {\rm h.c} \ .
\end{equation}
This term is crucial for generating a calculable Dirac mass term at one
loop level (see Fig~\ref{figure1}).

\vskip -0.1 truecm
The induced Yukawa coupling is given by:
\begin{equation}
\lambda_{\alpha\beta} =
\frac{1}{64\pi^2}g_{\alpha\gamma}\frac{m_{\gamma}}{v}f_{\gamma\beta}\sin(2\theta)
\ln\left(\frac{M_{S_2}}{M_{S_1}}\right)
\end{equation}
where $m_{\gamma}$ are the charged lepton masses, $v$ is the vacuum expectation value (VEV) of
the standard model higgs doublet, $M_{S_1}$ and $M_{S_2}$ are the masses of the
physical charged higgs fields, and $\sin(2\theta)$ is the mixing angle
between $S_1$ and $S_2$ given by:
\begin{equation}
\sin(2\theta) = \frac{2\kappa}{\sqrt{4\kappa^2 + (M_{S_2} - M_{S_1})}}
\end{equation}

Note that the Yukawa couplings in this model are naturally
small. Without any fine tuning of the parameters $f_{\alpha\beta}$,
$g_{\alpha\beta}$ it is simple to obtain $\lambda \sim 10^{-5}$. For right-handed neutrino
masses $M_{R_{\alpha}} \sim
10$ TeV note also that Majorana masses for the left-handed neutrinos
$m_{\nu} \leq 0.1$ eV are generated
via the seesaw mechanism. A detailed analysis of the neutrino masses and mixing
in this model can be found in ref.~\cite{Nasri:2001ax}.

Now let us turn to the issue of baryogenesis. As we mentioned above,
the baryon asymmetry may be generated by a lepton asymmetry. The first step in demonstrating
how this occurs is to identify the lepton number violating decays. It is usual to consider
the decay of the heavy right handed neutrinos $N_{R_{\alpha}}$ or of at least
two heavy higgs triplets (denoted by $\Delta_{1,2}$).
\begin{figure}[h]
\epsfxsize = 0.85 \hsize \epsfbox{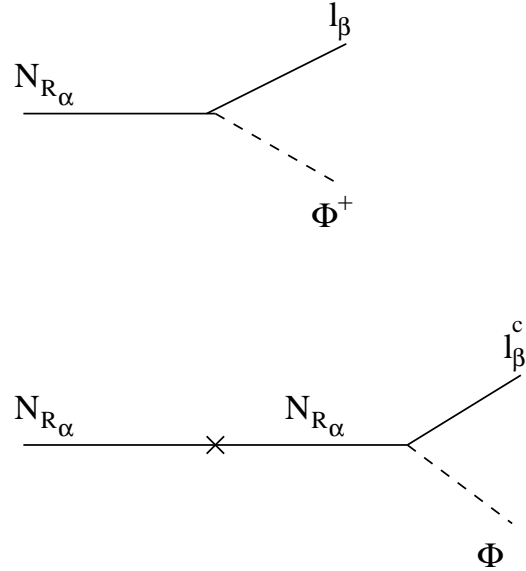}
\caption{\label{figure2}
Vertex and self energy diagrams for $N_R$ decay.
}
\end{figure}

\vskip -0.1 truecm
In most leptogenesis models
$N_{R_{\alpha}}$ and $\Delta_{1,2}$ are constrained to be heavy in order to suppress the masses
of the physical neutrinos, since the Dirac neutrino masses are of the order
charged lepton or quark masses. In our model the situation is very
different. Here the Yukawa couplings between the left-handed and the
right-handed neutrinos vanish at tree level, and are instead induced at one
loop. This ensures that they are naturally small and as a result
$N_{R_{\alpha}}$ need not be particularly heavy.

The relevant decays are (see fig~\ref{figure2})
\begin{eqnarray}
N_{R_{\alpha}}  &\longrightarrow &  l_{\alpha_L} + \overline \Phi \nonumber \\
N_{R_{\alpha}}  &\longrightarrow &   l_{\alpha_L}^C + \Phi \ \ ,
\label{decays}
\end{eqnarray}
which lead to a violation of lepton number by two units.

We now move on to the origin of CP violation in our model. The
couplings $f_{\alpha\beta}$ can
be made real by making a phase redefinition on the left handed fields. This is the same
freedom that ensures CP-conservation in the leptonic sector of the Zee model \cite{zee}.
However in our model the couplings $g_{\alpha\beta}$ may be complex, and therefore
the last term in (\ref{interaction}) is generically a source of CP violation in the
leptonic sector. Despite this, it is well-known that CPT-invariance implies that the
asymmetry in the decay rates vanishes at tree level.
\begin{figure}
\epsfxsize = 0.85 \hsize \epsfbox{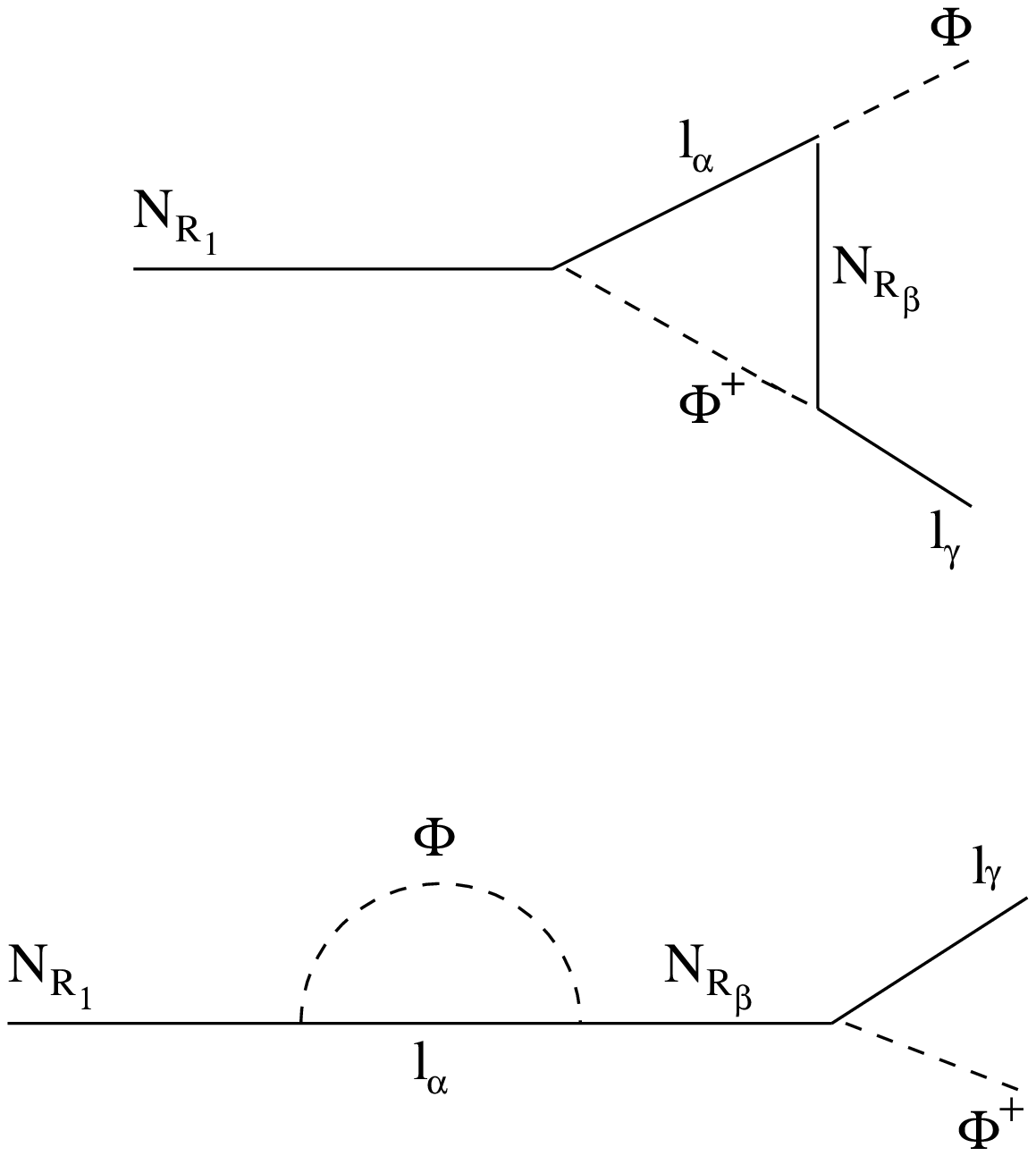}
\caption{\label{figure3}
Lepton number violating decays. The right-handed neutrino decays to leptons and antileptons.
}
\end{figure}

\vskip -0.1 truecm
Nevertheless, a non-zero asymmetry is
generated through  the
interference between the tree level diagram and the one loop
diagrams. Let us first consider the vertex correction and the wave
function renormalization of $N_{R_1}$ as in fig~(\ref{figure3}). The CP asymmetry
generated from the interference of these diagrams may be estimated as
\begin{equation}
\epsilon \simeq \lambda^2 \left(\frac{M_{R_1}}{M_{R_2}}\right) \ ,
\end{equation}
where $M_{R_1}$ is the mass of the lightest right-handed neutrino, and $M_{R_2}$
is the mass of the next heaviest. In this case the lepton asymmetry generated is
\begin{equation}
\eta_L < 10^{-13}\left(\frac{\lambda}{10^{-5}}\right)^2 \ ,
\end{equation}
which implies that the baryon asymmetry of the universe is highly suppressed for $\lambda \simeq 10^{-5}$.

However, consider the self-energy diagram in which, instead of the standard model
higgs field and the left-handed lepton, we have the charged higgs singlet $S_2$. In this case
the right-handed charged lepton is the most dominant contribution in
the interference with the tree level diagram because the
couplings $g_{\alpha\beta}$ can be much larger than the Yukawa
couplings $\lambda_{\alpha\beta}$.
\begin{figure}
\epsfxsize = 0.85 \hsize \epsfbox{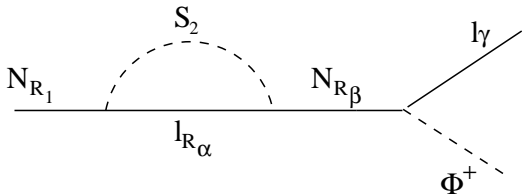}
\caption{\label{figure4}
Self energy diagram due the charged higgs singlet.
}
\end{figure}

\vskip -0.1 truecm

Let us assume that the the Yukawa couplings
in (\ref{interaction}) are all of the same order, and that the two right handed neutrinos are
almost degenerate. The CP asymmetry generated from the
decays (\ref{decays})
results from the interference between the tree level diagrams and the one loop diagrams
and is given by
\begin{equation}
\epsilon \simeq \frac{1}{16\pi}g^2xy^2\sin\delta \ ,
\end{equation}
where g is the Yukawa coupling of the right handed neutrinos with the
charged singlet higgs $S_2$ and $\delta =\arg(g^2)$. The
parameters $x$ and $y$ are defined by:
\begin{eqnarray}
x & = & \frac{M_{R_1}}{M_{R_2} - M_{R_1}}  \nonumber \\
y & = & 1 - \left(\frac{M_{S_2}}{M_{R_1}}\right)^2
\end{eqnarray}
Note that the CP asymmetry gets enhanced when the right handed
neutrinos are almost degenerate \cite{flanz}. Of course if they are exactly
degenerate the CP asymmetry vanishes, because in this case the relative
phase between the degenerate states can be rotated away. This is why
we need at least two non-degenerate right handed neutrino states. The
same considerations apply for leptogenesis via the decay of the higgs
triplet \cite{ma}.

The parameter $y$ can lead to a suppression when $M_{S_2} \simeq
M_{R_1}$. Indeed one needs y to be tiny in order to suppress the decay
$S_{2} \longrightarrow l_{R_{\alpha}} + N_{R_{\alpha}}$, otherwise it will be rapid enough to
wash out the baryon asymmetry generated at $T \simeq M_{R_1}$.

Now consider the required departure from thermal equilibrium in this model. The relevant
quantity parameterizing this is the ratio
\begin{equation}
K = \frac{\Gamma_{N_R}}{H} \ ,
\end{equation}
where $\Gamma_{N_R}$ is the decay rate of $N_{R_1}$, and
$H = \sqrt{1.7}g_{*}T^2/M_P$ is the expansion rate of the universe,
with $M_P\simeq 10^{19} GeV$ the Planck mass. If $K << 1$, then the
system is far from equilibrium and the lepton asymmetry will be $\eta_L
\simeq \frac{\epsilon}{g_{*}}$. However if $K >> 1$, then the baryon
asymmetry will be suppressed by a factor \cite{kolbandturner},
\begin{equation}
D \simeq \left\{\begin{array}{ll}
\sqrt{0.1K} \exp\left[-{\frac{4}{3}}(0.1K)^{\frac{1}{4}}\right] & \mbox{for $K\geq 10^6$} \\
\frac{1}{K(\ln K)^{0.6}} & \mbox{for $10<K\leq 10^6$} \\
\frac{1}{2K} & \mbox{for $1<K<10$}
\end{array}\right. \ .
\end{equation}
In our case
\begin{equation}
K = \frac{\lambda\lambda^{+}}{16\pi}\frac{M_P}{1.7\sqrt{g_{*}}M}
\end{equation}
and for $M_{R_1} \sim 10 TeV$ the dilution factor is $D \sim
3\times 10^{-3}$. Taking this into account, the baryon asymmetry generated by our model may
be estimated as
\begin{equation}
\eta \simeq 10^{-6}xyg^2\sin\delta \ .
\end{equation}
Although we may choose $x$ to be as large as $10^5$, over a wide
range of Yukawa couplings $f_{\alpha\beta}$ and $g_{\alpha\beta}$
reasonable values of the parameters involved are $x\sim 10^{4}$,
$y \sim 5.10^{-5}$ and $g^2\sin\delta \sim 1$. Using these values
we obtain $\eta \sim 10^{-11} - 10^{-10}$, which is in the range
required for successful primordial nucleosynthesis.

\begin{figure}
\epsfxsize = 0.85 \hsize \epsfbox{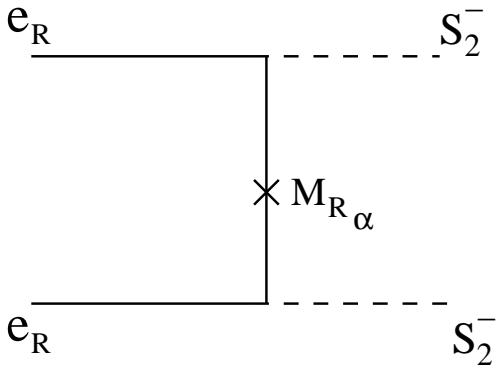}
\caption{\label{figure5}
The observable process $e_R +e_R \longrightarrow S_2^- + S_2^-$, predicted by the model.
}
\end{figure}

Now, as we have advertised, a very attractive feature of this model is its accessibility to
experimental tests. As a specific example, consider
the scattering of polarized beams of right-handed electrons. One observable
outcome of this would be the production of charged singlet fields $S_2^-$, mediated by the
right-handed neutrinos
\begin{equation}
e_R +e_R \longrightarrow S_2^- + S_2^- \ .
\end{equation}
This prediction depends crucially on how low the mass of the right-handed neutrino is
allowed to be in this model, since the Majorana mass insertion appears explicitly in the
relevant Feynman diagram (see fig~\ref{figure5}).

\vskip -0.1 truecm

In this letter we have described how leptogenesis, and hence baryogenesis, can occur quite
naturally in an existing particle physics model. The model involves a modest extension of the
standard electroweak theory, required to generate naturally low Dirac masses for the
left-handed neutrinos. In
this sense, the generation of the baryon asymmetry of the universe is natural in this model.
Leptogenesis is traditionally thought to occur at the relatively high scale of $10^{10}$ GeV. However,
a very attractive feature of the model presented here is that only scales of less than or of
order $10$ TeV are required. This means that it is feasible to test this model in collider
experiments, perhaps at a high-center of mass energy $e^+,e^-$ machine.

The authors would like to thank T. Hambye for useful remarks on
and an important numerical correction to an earlier version of the
manuscript. This work was supported by NSF grant PHY-0094122 to
MT, and by the DOE (SN).

\end{document}